\documentclass[]{aipproc}

\layoutstyle{6x9}

\SetInternalRegister\hbadness{8000} 

\usepackage{epsfig}
\usepackage[scriptsize]{subfigure} 

\begin{document}

\newcommand{\reaction}{\mbox{$pp\rightarrow\,pp\eta$ }}
\newcommand{\reaktion}{\mbox{${\vec{p}}p\rightarrow\,pp\eta$ }}

\title 
      [Analysis of the ${\eta}$ meson production mechanism]
      {Analysis of the ${\eta}$ meson production mechanism via the ${\vec{p}p\rightarrow pp\eta}$ reaction}


\newcommand{\ikpjuel}{IKP, Forschungszentrum J\"{u}lich, D-52425 J\"{u}lich, Germany}
\newcommand{\ikpmue}{IKP, Westf\"{a}lische Wilhelms--Universit\"{a}t, D-48149 M\"{u}nster, Germany}
\newcommand{\cracow}{M.~Smoluchowski Institute of Physics, Jagellonian University, PL-30-059 Cracow, Poland}
\newcommand{\nphycracow}{H. Niewodnicza{\,n}ski Institute of Nuclear Physics, PL-31-342 Cracow, Poland}
\newcommand{\catowice}{Institute of Physics, University of Silesia, PL-40-007 Katowice, Poland}
\newcommand{\zeljuel}{ZEL,  Forschungszentrum J\"{u}lich, D-52425 J\"{u}lich,  Germany}

\author{R.~Czy{\.{z}}ykiewicz}{address={\cracow,\ikpjuel}}
\author{H.-H.~Adam}{address={\ikpmue}}
\author{A.~Budzanowski}{address={\nphycracow}}
\author{D.~Grzonka}{address={\ikpjuel}}
\author{M.~Janusz}{address={\cracow}}
\author{L.~Jarczyk}{address={\cracow}}
\author{B.~Kamys}{address={\cracow}}
\author{A.~Khoukaz}{address={\ikpmue}}
\author{K.~Kilian}{address={\ikpjuel}}
\author{P.~Kowina}{address={\ikpjuel}}
\author{P.~Moskal}{address={\ikpjuel}}
\author{W.~Oelert}{address={\ikpjuel}}
\author{C.~Piskor-Ignatowicz}{address={\cracow}}
\author{J.~Przerwa}{address={\cracow}}
\author{T.~Ro{\.{z}}ek}{address={\ikpjuel}}
\author{R.~Santo}{address={\ikpmue}}
\author{G.~Schepers}{address={\ikpjuel}}
\author{T.~Sefzick}{address={\ikpjuel}}
\author{M.~Siemaszko}{address={\catowice}}
\author{J.~Smyrski}{address={\cracow}}
\author{A.~T\"aschner}{address={\ikpmue}}
\author{P.~Winter}{address={\ikpjuel}}
\author{M.~Wolke}{address={\ikpjuel}}
\author{P.~W{\"u}stner}{address={\zeljuel}}
\author{W.~Zipper}{address={\catowice}}
 \copyrightholder{Acoustical Scociety of America}
\copyrightyear  {2001}

\begin{abstract}
Polarisation observables constitute a powerful tool for 
establishing the production mechanism of the $\eta$ meson
and for infering the presence of higher partial 
waves in the final system. 
Measurements of the proton analysing power for the \reaktion reaction have been 
performed by the COSY-11 group at three different excess energies:
Q=10, 37 and 40 MeV. Data at Q=40 MeV indicate that 
the $\eta$ meson is probably produced in partial waves 
higher than $s$ wave.
\end{abstract}

\date{\today}

\maketitle

\section{Introduction}

	Despite the fact that the discovery of the $\eta$ meson took place over fourty years ago
\cite{pevsner:61}, its production mechanism still remains an open question.
Based on the close-to-threshold total cross section measurements for the \reaction reaction
\cite{bergdolt:93,chiavassa:94,calen:96,calen:97,hibou:98,smyrski:00}, 
investigations on differential cross sections for this reaction
\cite{calen:99,tatischeff:00,moskal:01-2,moskal:02,moskal:02-2,abdelbary:02,moskal:02-3,moskal:03}
and recently performed measurements of the proton analysing power for the
\reaktion reaction~\cite{winter:02-2,winter:02-1-en},   
there is a consensus that in the NN collisions the $\eta$ meson is produced
in a two-step process, where in the first stage exchange of one of the pseudoscalar or vector 
mesons excites the $S_{11}$ resonance 
and subsequently this resonance decays into a N$\eta$ pair. 
The $S_{11}(1535)$ resonance seems to play an important role 
as an intermediate state since it has a large width covering 
the threshold energy for the \reaction reaction, and it couples
strongly to the N$\eta$ system with a branching ratio 
corresponding to 30-55 \% ~\cite{caso:98}.

\begin{figure}[h]
\epsfig{file=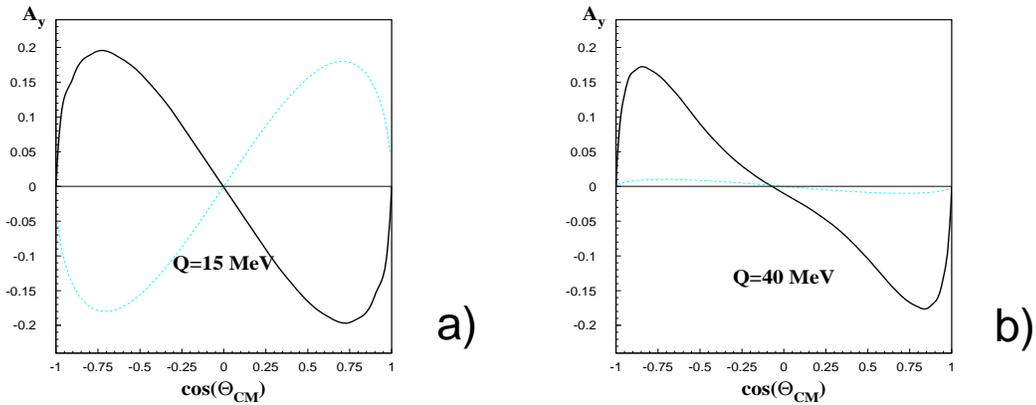,scale=0.7}
\caption{Predictions of the angular dependence for proton analysing power
at (a) Q=15 MeV and (b) Q=40 MeV. 
The solid lines
are the predictions of reference \cite{nakayama:02},
whereas the dotted lines represent the 
calculations according to reference \cite{faldt:01}.  
$\theta_{CM}$ is the centre-of-mass polar angle of the $\eta$ meson.
\label{wilkin_nakayama}}
\end{figure}

	Although there exists a variety of models based on different
assumptions concerning the mechanism of the $\eta$ production, these models are in 
quite good agreement with the existing data on close-to-threshold 
total cross sections for the \reaction reaction~\cite{moskal:02-4}. 
As it was shown in reference~\cite{nakayama:02},
the excitation function for the \reaction process can be equally well 
described by the intermediate excitation of the S$_{11}$ via the exchange of either pseudoscalar  
or vector mesons. 
This implies that more limitations have to be added to the 
models in order to extract the way the $\eta$ meson is really being created. 
One solution would be the verification of different models by means of the polarisation observables. 
At present there exist two models that predict 
the energy dependence of the proton analysing power for the \reaktion reaction
\cite{nakayama:02,faldt:01}.      
Theoretical forecasts for the two different excess energies Q=15 MeV and Q=40 MeV 
are shown in fig.\ref{wilkin_nakayama}.  
There are significant differences between the models   
visible in both: the relative sign (at lower Q values) and in the magnitude of the 
proton analysing power. Measurements of this observable 
might therefore help to establish the adequacy of mechanisms of 
$\eta$ meson production.

\section{Experiments}

	There have been, so far, performed three measurements of 
proton analysing power for the \reaktion reaction at 
different excess energies. All these experiments made use of the  
COSY-11 installation~\cite{brauksiepe:96}, which is an internal detection setup 
mounted inside the COSY ring\footnote{Accelerator's name 'COSY' is an acronym of  
COoler SYnchrotron.}~\cite{maier:97nim} in the Research Centre J{\"u}lich.  
Experiments were performed in three separate runs in January '01 (at excess 
energy Q=40 MeV), September '02 (Q=37 MeV) and April '03 (Q=10 MeV).
Data taken during the first run have been analysed and  
published~\cite{winter:02-2} whereas the data from the two last runs  
await the full analysis. In the following the results from measurement
at the excess energy Q=40 MeV will be presented as well as some preliminary 
estimations for the other runs.

\begin{figure}[h]
\epsfig{file=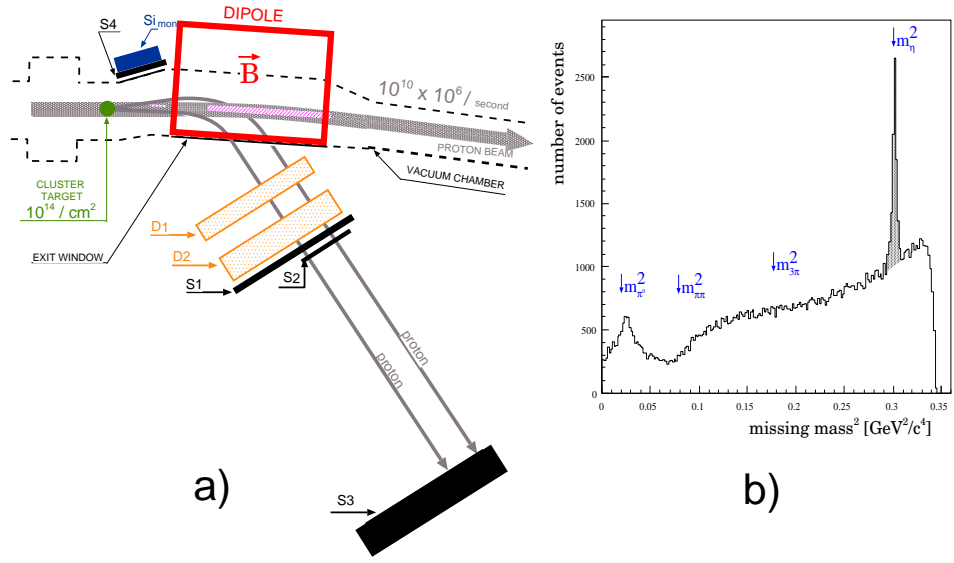,scale=0.6}
\caption{ (a) Schematic view of the COSY-11 detection setup. D1 and D2 
denote the two stacks of drift chambers with 6 and 8 detection planes 
respectively. S1, S2, S3 and S4 are the scintillator detectors. The time 
of flight is measured between the S1 scintillator hodoscope and the S3 scintillation wall.
Having known the velocity of the particle from the time of 
flight measurement and the particle's momentum from the track reconstruction in the magnetic 
field inside the dipole magnet one is able to reconstruct the particle's mass. Si$_{mon}$
are the silicon detectors used for the detection of elastically scattered protons.
For more details the reader is refered to~\cite{brauksiepe:96,moskal:01}. 
(b) Square of the missing mass
for events with two protons in the final state. 
Values indicated by arrows are the literature masses~\cite{groom:00} of individual particles or
systems of particles. Figure is adapted from~\cite{winter:02-1-en}.} 
\label{COSY11} 
\end{figure}

	The mentioned COSY-11 detection setup is displayed 
schematically in fig.~\ref{COSY11}a.
Due to the short lifetime, the $\eta$ meson cannot be registered in any 
of the detectors available at COSY-11. Its identification, however, can be performed 
using the missing mass method. Having known the beam and target's four-momenta, denoted
as $P_{beam}$ and $P_{target}$ respectively, and having reconstructed the four-momenta of 
the two protons in the exit channel~($P_1$ and $P_2$), it is possible to determine the square mass of the 
unregistered particle or system of particles according to the formula:

\begin{equation}
m{_X}^{2} = ( P_{beam} + P_{target} - P_1 - P_2 )^{2}.
\end{equation}  

The missing mass spectrum for events with two protons in the exit channel, 
obtained from the whole measurement at Q=40 MeV is shown in 
figure~\ref{COSY11}b. A high peak at the value of the $\eta$ mass square is clearly seen.
with about 6000 events inside the peak above the background. The background
mainly originates from multi-pion production.  
The increase of the background in the higher-energy part of the spectrum 
is a reflection of the increasing acceptance of the detection system in this region.
The mass resolution achieved during this measurement was equal to $\sigma_{m_{\eta}}=1.6$~MeV/c$^{2}$.

	In order to determine the analysing power\footnote{For 
detailed calculation of the analysing power, which takes into account the 
efficiency corrections, the reader is referred to~\cite{winter:02-2}.} 
one has at the same time to monitor both the luminosity
and the beam polarisation. 
During the first two measurements we made use of the EDDA experimental setup~\cite{bisplinghoff:93}
in order to extract the beam polarisation, whereas in the last two 
runs we monitored the polarisation by means of 
the COSY internal polarimeter~\cite{bauer:01} 
and our own monitoring system, allowing to measure elastically 
scattered protons in the horizontal and vertical planes. 


\section{Results}
\begin{figure}[h]
\epsfig{file=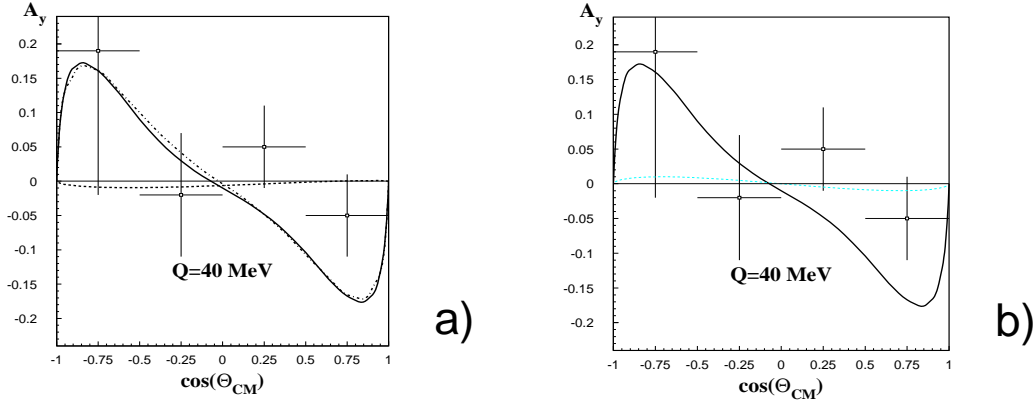,scale=0.7}
\caption{ (a) Analysing power for the \reaktion reaction at Q=40 MeV. Curves in this figure are adapted 
from~\cite{nakayama:03}.
(b) Comparison of experimental data with predictions of different models. The labelling of curves has been 
explained in the caption of figure~\ref{wilkin_nakayama}. \label{analysing}}
\end{figure}

The analysing power values for the \reaktion reaction 
as obtained from the measurement at an excess energy
Q=40 are presented in figure~\ref{analysing}. The range of the $\eta$
centre-of-mass polar angle has
been divided into four bins, with a width of 0.5 in cos$(\theta_{CM})$. There are about
1000-1500 events in each bin.
Figure~\ref{analysing}a shows the results of analysis within the framework of  
the vector meson exchange model~\cite{nakayama:03} 
for the \reaktion reaction. The dashed line in this figure represents 
the $\eta$ $s+p$ partial wave contribution, the $s+p+d$ waves are 
represented by the dashed-dotted line, 
whereas the full model calculations,
taking into account also the higer partial waves are 
shown as a solid line, which is close to 
the dash-dotted line. It is worth to note that the pure 
$s$ wave should force A$_y$ to be equal to 0.
However,
at present level of the measurement's accuracy we are not able to 
infer any quantitative conclusions. 

	The interference term between Pp and Ps waves (denoted as G$_1^{y0}$,
where we follow the notation from~\cite{meyer:01}) as well as 
the sum of (Pp)$^2$ and (SsSd) interference terms (H$_1^{y0}$ + I$_1^{y0}$) 
have been calculated and 
are equal to~\cite{winter:02-2}:
\begin{center}
 G$_1^{y0}$ = ( 0.003 $\pm$ 0.004 ) ${\mu}$b, \\
\end{center}
\begin{center}
 H$_1^{y0}$ + I$_1^{y0}$ = ( - 0.005 $\pm$ 0.005 ) ${\mu}$b.
\end{center}
The vanishing value of G$_1^{y0}$ may suggest that there is no interference between Pp and Ps waves in 
the final state system. Unfortunatelly, during the first measurement the precision of data was 
not good enough to disentangle the sum of H$_1^{y0}$ and I$_1^{y0}$.

\section{Outlook}
	As can be seen from figure~\ref{analysing}b 
both model predictions lie within 2$\sigma$ distance from experimental data, therefore
at the level of accuracy
obtained during the first measurement we are not able to distinguish 
between two different hypotheses of the $\eta$ production. 
Thus further investigations were necessary.  
During the last two measurements 
the beam polarisation was increased from about 50\% (which 
was the averaged polarisation during the first run) up to 75-80\%. 
This nice feature, together with the 
fact that the luminosity integrated over the whole measurement's period was larger 
by a factor 1.5, should result in about two times better 
accuracy of the last two measurements. 
Moreover, the difference between predictions of two mentioned models are 
the largest at the excess energy Q=10 MeV, studied in present experiments,
which should additionally help to discriminate one of the reactions models.   

\bibliographystyle{aipproc}
\bibliography{abbrev,general}

\end{document}